\documentclass[aps,prl,twocolumn]{revtex4}
\usepackage{amsmath,amssymb}
\input epsf
\usepackage{verbatim}
\usepackage{graphicx}
\usepackage{bm}
\usepackage{natbib}
\def\pmb#1{\setbox0=\hbox{#1}
\kern-.025em\copy0\kern-\wd0 \kern-.05em\copy0\kern-\wd0
\kern-.025em\raise.0433em\box0}

\newcommand{\beq}{\begin{equation}}
\newcommand{\eeq}{\end{equation}}
\newcommand{\ba}{\begin{eqnarray}}
\newcommand{\ea}{\end{eqnarray}}

\input epsf

\usepackage[english]{babel}
\begin{document}
%\doublespace

\title[]{Moulding and shielding flexural waves in elastic plates
% lying atop a
 % Faqir's bed
 % of nails
}
\author{T. Antonakakis$^{1,2}$, R.~V. Craster $^2$, S. Guenneau$^3$}
\affiliation{$^1$ European Organization for Nuclear Research, CERN CH-1211, Geneva 23,
Switzerland }
\affiliation{$^2$ Department of Mathematics, Imperial College London, London SW7 2AZ, UK }
\affiliation{$^3$ Institut Fresnel, UMR CNRS 6133, University of Aix$-$Marseille, Marseille, France }

\begin{abstract}
  Platonic crystals (PCs) are the elastic plate analogue of the
  photonic crystals widely used in optics, and are thin structured
  elastic plates along which flexural waves cannot propagate within
  certain stop band frequency intervals.  The practical importance of
  PCs is twofold: these can be used either in the design of
  microstructured acoustic metamaterials or as an approximate model
  for surface elastic waves propagating in meter scale seismic
  metamaterials.  Here, we make use of the band spectrum of PCs
  created by an array of very small clamped circles to 
  achieve surface wave reflectors at very large wavelengths, a flat
  lens, an endoscope, a directive antenna near stop band frequencies
  and cloaking from Dirac-like cones. The limit as the circles reduce to
  points is
  particularly appealing as there is an exact dispersion relation
  available so the origin of these phenomena can be explained and
  interpreted using Fourier series and high frequency homogenization.
%These
%techniques reveal there is a countable set of generalized
%Dirac cones (i.e. any intersection of an even number of
%linear curves with a quadratic curve on
%dispersion diagrams) for pinned PCs, leading to near-zero index cloaking
%at many (isolated) frequencies of surface elastic waves.

\end{abstract}
%\label{firstpage}
\maketitle

There has been much interest over the past 20 years in the analysis of
elastic waves in thin 
plates in the continuum mechanics community
\cite{howe94a,murphy94a,norris95a,mace96a,abrahams00a,movchan07c,antonakakis12a};
an interest
renewed in the metamaterial community with the theoretical proposals \cite{farhat09a,farhat10a,farhat10b,khelif12},
and their subsequent experimental validation \cite{bonello10,stenger12,veres12},
of broadband cloaks and negatively refracting flat lenses for flexural waves.
%The theory describing the flexural motion of thin homogeneous plates
%is well established and can be found in many
%classical books
%, one which is well-known was published 55 years ago
%\cite{viktorov67a,graff75a,landau70a}. 

One of the attractions of platonics is that 
 much of the physics of photonic crystals can be translated into platonic crystals
(PCs). %But 
There are %some 
 mathematical subtleties in the analysis, and
numerics, of
the scattering
of flexural waves \cite{norris95a} owing to the fourth-order
derivatives, versus the usual second-order derivatives for the wave
equation of optics, 
 involved 
in the governing equations; even the waves within a perfect plate
have differences from those of the wave equation as they are not
dispersionless. 
% Thus one
%requires more involved numerical treatments to take into account additional
%imit conditions compared with second order acoustic wave equations.

There is a long history of wave propagation along periodically
supported plates motivated by ribbed structures in underwater acoustics,
\cite{mead96a}; the periodically constrained plate \cite{mace96a} being 
considered in detail. 
%In 2007, Evans and Porter \cite{evans07a} presented the first
%analysis of the propagation of bending waves in a doubly
%periodic array of rigid pins in an infinite thin plane, the object of the present
%study. 
 Movchan et al. \cite{movchan07c} also give a complementary treatment
% detailed analysis
%of the band structure Floquet-Bloch spectral problems
 for the flexural
plate containing square arrays of circular holes or clamped
circles 
% with either clamped or stress free boundary conditions,
 using
lattice sums and multipoles with 
% expansions combined with lattice sums.
a dilute
%composite 
 limit giving approximate dispersion relations %appeared 
 in closed form.
%therein.
 Fourier series expansions introduced
in \cite{mace96a,evans07a} and %further
  applied in
\cite{antonakakis12a} also allow for a
highly accurate analytic expression for the exact dispersion equation
for plates when the radius of the clamped circles tends to zero and
the clamping is at points. We follow this latter route, which combined with
the recently developed high
frequency homogenization theory (HFH) \cite{craster10a},
uncovers the physics of point clamped 
%\footnote{The well-known circus
%  trick, or physics demonstration, of a person lying atop a bed of
% nails being traditionally attributed to travelling Faqirs}
plates; a 
striking illustration %of HFH
(lensing) is shown in Fig. \ref{fig1}(b), which
contrasts with a shielding effect at a lower
frequency in Fig. \ref{fig1}(a). The latter
obviously has potential application in
seismic metamaterials \cite{menard12}
for anti-earthquake systems.

Much has been said about control of light \cite{sar_rpp05}, sound,
water, or shear (SH) waves \cite{thebook} using the rich 
behaviour encapsulated by the  dispersion curves
of bi-periodic structures, modelled by a Helmholtz equation, up to minor changes in the
normalization of material parameters, and choice of boundary conditions (e.g. Dirichlet or
Neumann for clamped or freely vibrating inclusions in the context of SH waves).
However, when one moves
into the area of elastic waves, governed by Navier equations, it is no longer possible to reduce the analysis to a single scalar partial differential equation (PDE), as shear and pressure waves do
couple at boundaries. There is nevertheless, the simplified framework of the
Kirchhoff-Love plate theory \cite{graff75a} that allows for bending
moments and transverse shear forces to be taken into account via a fourth-order PDE
for the out-of-plane plate displacement field.
This plate theory is a natural extension of the Helmholtz equation to a generic model for
flexural wave propagation through any spatially varying thin elastic medium.
It offers a very convenient mathematical model for any physicist
wishing to grasp (some of) the physics of PCs using earlier
knowledge in photonic or phononic crystals.
However, while the Helmholtz equation can, with appropriate notational and
linguistic changes, hold for acoustic, electromagnetic, water or
out-of-plane elastic waves and so encompasses many possible 
applications, the Kirchhoff-Love plate
theory is dedicated to the analysis of flexural waves. For
instance, it says little about propagation of in-plane
elastic waves in platonic crystals.

Flexural, also called bending, waves are fundamentally different in
character from compressional acoustic or electromagnetic waves; they
 model the lowest frequency waveguide mode for elastic
 waves and therefore are dispersive. For these low
 frequencies, equivalently thin plates, the phase and group velocities
 scale with square root of the frequency and this is reflected in the model equations
 which are fourth order in space and second order in time. A reduced
 model is derived in \cite{landau70a,graff75a} for flexural waves, 
%This simple relationship is a
% consequence of the stiffness/thickness relationship for thin plates
% in bending.
 and in the frequency domain it is  
\beq
\left(\frac{\partial^2}{\partial x_1^2}+\frac{\partial^2}{\partial x_2^2}+\Omega\right)
\left(\frac{\partial^2}{\partial x_1^2}+\frac{\partial^2}{\partial x_2^2}- \Omega\right)u=0,
\label{eq:helmholtz}
\eeq
% \cite{landau70a,graff75a}  
for $u(x_1,x_2)$ 
 on the square cell $-1<x_1,x_2<1$ where this has been
 non-dimensionalised: All results presented are in non-dimensional
 units and one can return to the dimensional setting using that $\Omega^2=12(1-\nu^2)\rho \omega^2/(Eh^2)$, where $\rho$, $h$, $E$, $\nu$ are
density, thickness, Young's modulus and Poisson's ratio of the plate, respectively, and $\omega$ is the angular wave
frequency; the plate contains a square array of very small
clamped circles, so both $u=0$ and $\partial u/\partial r=0$ on, and within, a radius $r=a$ centred
at each lattice point of the array. 

In the context of optics, (\ref{eq:helmholtz}) reduces to
a second order PDE where only the first factor of (\ref{eq:helmholtz})
appears, in which case Dirichlet boundary conditions
are a good model for infinite conducting thin wires at GHz
frequencies, with the unknown $u$ of such a Helmholtz
equation being the longitudinal component of the electric  field $E_z$
in TM polarization and the spectral parameter $\Omega$ is associated with
$\omega^2 \varepsilon(x_1,x_2)/c^2$ whereby $\omega$ is the
electromagnetic wave frequency, $\varepsilon$ is the relative permittivity
and $c$ is the speed of light in a vacuum. Such a model has been known
since the mid 90s to lead to a zero frequency stop band associated
with very low frequency plasmons \cite{nicorovici95a,pendry96a}.

\begin{figure}[h!]
\includegraphics[width=9cm]{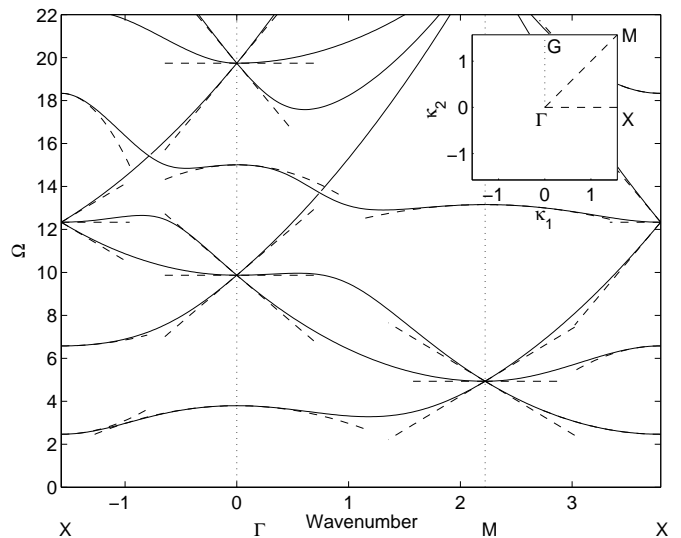}
\caption{Bloch dispersion curves around the edges
of the Brillouin zone $\Gamma XM$ (inset) for an array
of clamped circles (radius $0.01$) of pitch $2$.
One notes the stop band for
$\Omega\in[0,\pi^2/4]$ and three Dirac-like cones
at frequencies $\Omega=\pi^2/2$ ($X$ point),
$\pi^2$ and $2\pi^2$ ($\Gamma$
point). Solid curves are computed with FEM,
and dashed curves are from HFH.}
\label{fig2}
\end{figure}

\begin{figure}[h!]
\includegraphics[width=8.4cm]{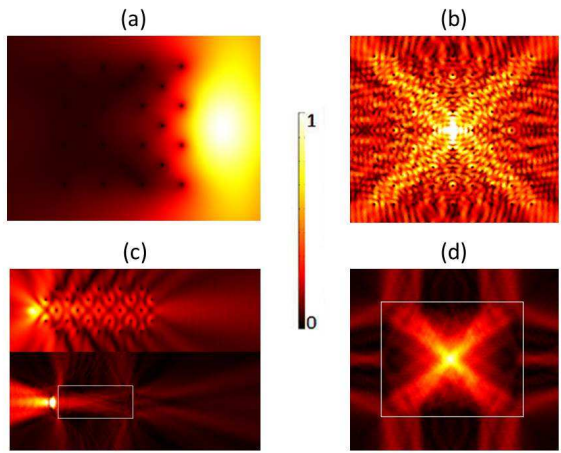}
\caption{(a) A forcing generates
a bending wave totally reflected by the array (with 22 clamped circles),
throughout the zero-frequency
stop band of Fig. \ref{fig2} (here $\Omega=2$). From
the four-fold symmetry
of the array and linearity of Eq.\ref{eq:helmholtz},
the defect at the center of the array (one missing
circle) is shielded from the bending
waves. (b) A point source placed in the centre of an array of $80$ circles of radius $0.01$ in a regular square
orientation at the same frequency of $\Omega=6.58$ produces a X
shape. (c) An endoscope effect of bending wave excited by a point
forcing of normalized frequency $\Omega=6.58$ through an array (pitch
$2$) of $27$ clamped circles 
(radius $0.01$) tilted through an angle $\pi/4$. (d) Represents the
same X shape effect produced by an effective material created by
Eq. (\ref{eq:f_0}) with $T_{11}T_{22}<0$ yielding characteristic type
of solutions. Panels (b) and (d) use the PC array in the usual $\Gamma
X$ symmetry direction whereas panels (a), (d) rotate the array to use
the $\Gamma M$ symmetry direction. 
%s placed at all four Cardinal points. 
}
\label{fig1}
\end{figure}

For waves propagating through an infinite PC containing a square array
of identical defects, one can
invoke Bloch's theorem \cite{kittel96a,brillouin53a} and simply
consider a square cell of sidelength $2$ with Bloch conditions applied to the
edges. %The quasi-periodic Bloch boundary conditions are
%\beq
%  u(1,x_2)=e^{2\ri\kappa_1}u(-1,x_2),
%\quad
 % u_{x_1}(1,x_2)=e^{2\ri\kappa_1}u_{x_1}(-1,x_2),
%\label{bloch1}
%\eeq
%\beq
 % u(x_1,1)=e^{2\ri\kappa_2}u(x_1,-1),
%\quad
 % u_{x_2}(x_1,1)=e^{2\ri\kappa_2}u_{x_2}(x_1,-1),
%\label{bloch2}
%\eeq
 The Bloch wave-vector $\bm\kappa=(\kappa_1,\kappa_2)$
 characterizes the phase-shift going from one cell to the
 next. Applying the clamped boundary conditions at the local origin at
 each lattice point, leads to an exact solution when the cylinder
 radius shrinks to a point, as 
readily found  \cite{mace96a,antonakakis12a}  which gives 
%as 
%\beq
% u({\bf x})= \exp(i{\bm \kappa}\cdot{\bf x})\sum_{n_1,n_2}
%  \frac{\exp(-i\pi{\bf N}\cdot{\bf x})}{[(\kappa_1-\pi
%    n_1)^2+(\kappa_2-\pi n_2)^2]^2-\Omega^2},
%\label{eq:2du}
%\eeq
% and enforcing the condition at the origin gives 
the dispersion
 relation $D(\bm\kappa,\Omega)=0$ where 
\beq
D(\bm\kappa,\Omega)=
\sum_{n_1,n_2=-\infty}^\infty
%\sum_{n_2=-\infty}^\infty
\frac{1}{[(\pi  n_1-\kappa_1)^2 +(\pi n_2 -\kappa_2)^2]^2-\Omega^2}.
\label{eq:2d_dispersion}
\eeq 
 As is well-known in solid state physics \cite{brillouin53a}
only a limited range of wavenumbers need normally be considered, namely
the wavenumbers along the right-angled triangle $\Gamma XM$ shown in
the irreducible Brillouin zone inset in Fig. \ref{fig2}. There
are however some exceptions,  where it is 
possible to miss important details such as the stop band minima/maxima not
occuring at the edges of the Brillouin zone
\cite{harrison07a,adams08a,adams08b}  and flat bands leading to strong
anisotropy and slow light \cite{craster12a}. 

%figure * goes across both columns

% In this two dimensional example a
 % Bloch wavenumber vector ${\bm\kappa}=(\kappa_1,\kappa_2)$ is used
 % and the dispersion relation can be characterised completely by
 % considering the irreducible Brillouin zone $\Gamma XM$.

\begin{figure}[h!]
\includegraphics[width=8.6cm]{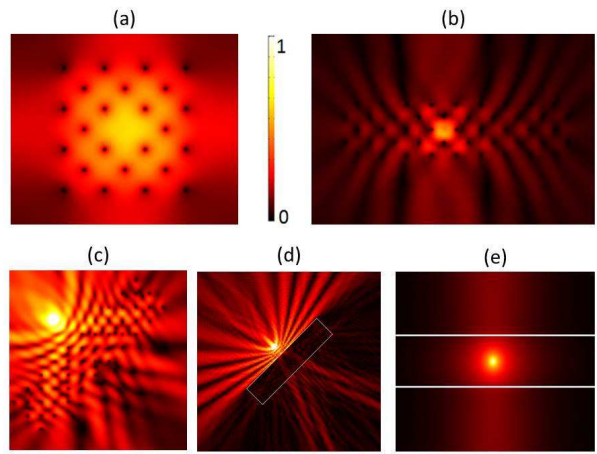}
\caption{A bending wave excited by a forcing of normalized frequency $\Omega=3.7952$
(that of vanishing group velocity at $\Gamma$ point in Fig. \ref{fig2})
inside an array of clamped circles (pitch $2$, radius $0.01$) tilted through an angle $\pi/4$ gives rise to
respectively four (a) and two (b) highly-directed beams outside $24$ circles making
a square (a) and $47$ circles making a rectangle (b). Focusing through a slab of $48$ circles is shown in (c) with its HFH equivalent in (d) for a point forcing at normalized frequency of $\Omega=6.58$. The focusing 
(d) and antenna (e) effects are simulations of the
continuum PDEs (\ref{eq:f_0}) generated 
by HFH with respective $(T_{11},T_{22})$ coefficients of $(25.65,-11.18)$ at point $X$ and $(6.2524,6.2524)$ at point $\Gamma$, where the effective
media are highlighted by white lines. Panels (a) and (b) use the PC array in the $\Gamma
M$ symmetry direction and panel (c) uses the $\Gamma M$ symmetry direction. }
\label{fig3}
\end{figure}

\begin{figure}[h!]
\includegraphics[width=8.6cm]{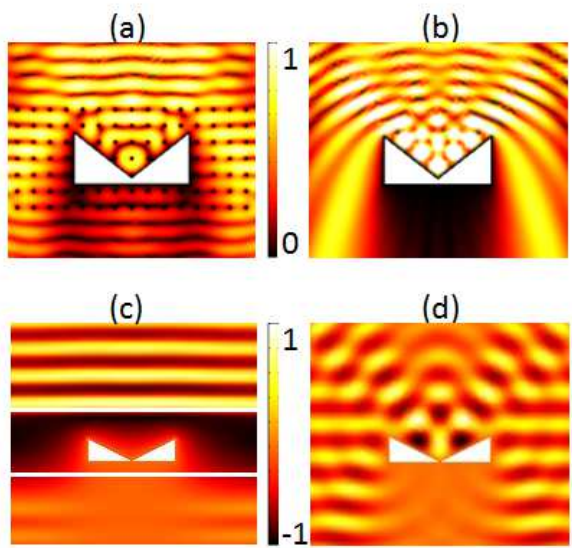}
\caption{A plane bending wave of normalized frequency $\Omega=9.7$
(just below the first Dirac-like cone at $\Gamma$ point in Fig. \ref{fig2})
incident from the top on an array of clamped circles
(pitch $2$, radius $0.01$)
undergoes considerably less scattering (a) than
by a clamped obstacle on its own (b); panels (a,b) are from finite
element simulation.
The asymptotic HFH PDEs from equation (\ref{eq:f_0Dirac}) capture the
essence of physics, and the equivalent results are shown for
cloaking (c) and scattering (d) by the same
clamped obstacle.}
\label{fig4}
\end{figure}  

\begin{figure}[h!]
\includegraphics[width=8.6cm]{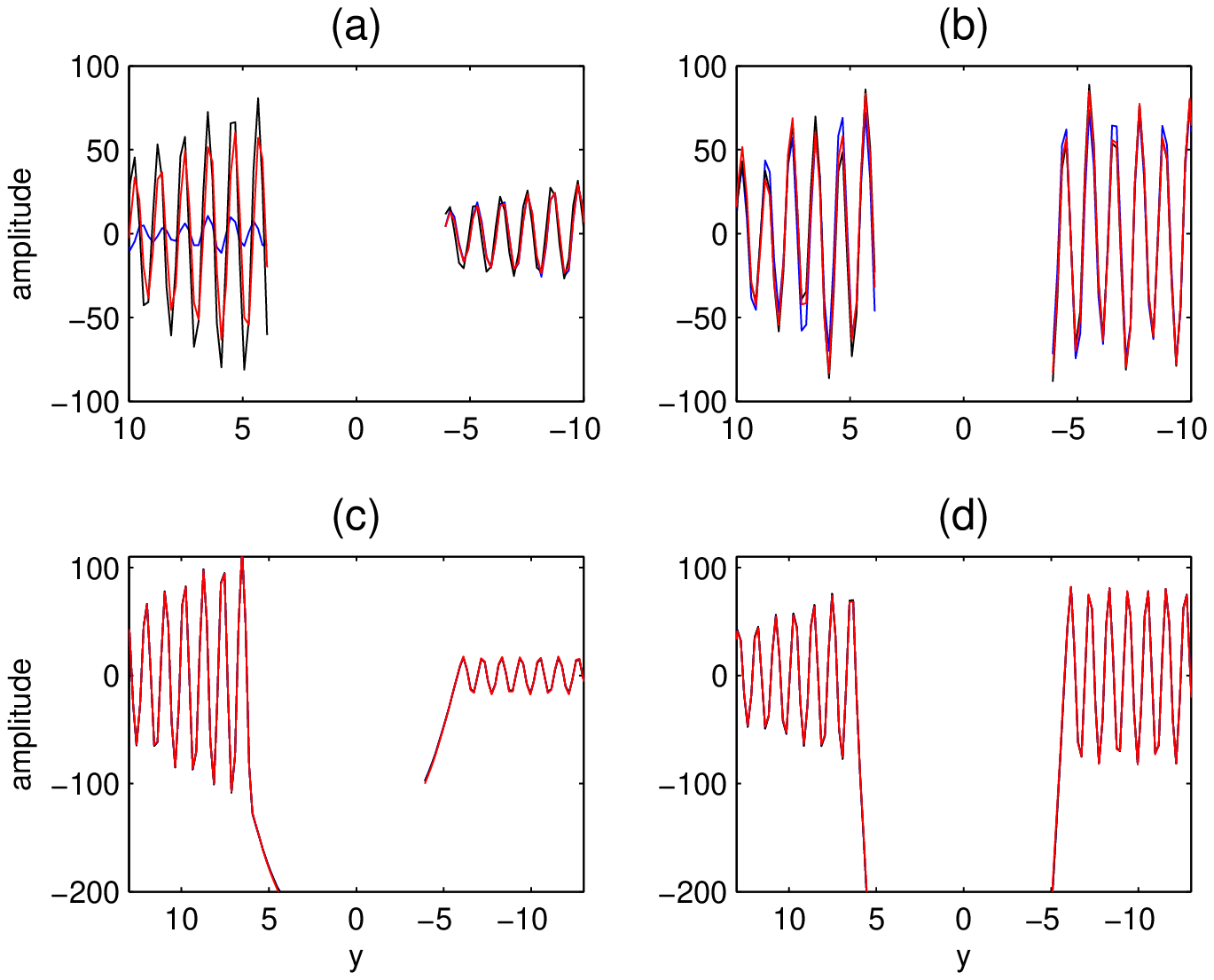}
\caption{Shape independence for three obstacles of same area, namely a square, $\pi/4$ tilted square and a circle in the same setting as in Fig. \ref{fig4}. (a) and (b) plot forward and back scattering on the respective paths $x=0$ and $x=15$ (outside the obstacle abscissa) for three different obstacle shapes without any periodic medium. (c) and (d) plot forward and back scattering on the same paths only this time with an effective medium that surrounds the obstacle just like in Fig. \ref{fig4}(c).}
\label{fig5}
\end{figure}  

 The dispersion diagram is shown in Fig. \ref{fig2}; the
 singularities of the summand in Eq. (\ref{eq:2d_dispersion})
 correspond to solutions within a square cell (without the clamped array) 
 satisfying the Bloch conditions at the edges, in some cases these
 singular solutions also satisfy the conditions at the support and are
 therefore true solutions to the problem.
%, a similar situation occurs
% in the clamped case considered using multipoles in
% \cite{movchan07c}. 
 Solid lines in Fig. \ref{fig2} are the dispersion curves
% that are branches of the dispersion relation 
 double checked by
the finite element method (FEM) using the commercial package
Comsol, notable features are
 the zero-frequency stop-band and also crossings of branches at the
 edges of the Brillouin zone. 
%Branches of the dispersion relation that
 %touch the edges of the Brillouin zone singly fall into two
 %categories, those with multiple modes emerging at a same standing wave frequency (such as
 %the lowest branch touching the left handside of the figure at M)
%and those that are completely alone (such as the second lowest branch on
 %the left at M).
 Importantly, HFH 
 \cite{antonakakis12a} 
captures the physics of this band diagram, with the 
asymptotic theory giving dashed curves that fit very well with the FEM
(solid curves) near the band-gap edges which are the zero-group velocity and Dirac-like
cones. %The theory of HFH for flexural waves in explains all the following equations in detail.
The multiple scales used for HFH are a short-scale $\xi_i=x_i/l$ and a
long-scale  
$X_i=x_i/L$ for $i=1,2$ where $l$ and $L$ represent respectively the 
characteristic small scale (half length of a cell) and the long scale. 
A small parameter is formed as $\epsilon=l/L$ and an expansion of
$\Omega^2=\Omega_0^2+\epsilon\Omega_1^2+\ldots $ and $u=u_0+\epsilon
u_1+\ldots$ 
is posed with respect to this parameter. Here $\Omega_0$ is a
standing wave frequency given at the points $X$, $\Gamma$ and $M$ and
a perturbation scheme can be developed about the band edges. The
leading order term for $u$ is $u_0(\boldsymbol\xi,{\bf X})=f_0({\bf
  X})U_0(\boldsymbol\xi)$ where the function $f_0$ representing an
envelope of the solution $u$ is obtained by an equation only on the
long scale: The theory is detailed in  \cite{antonakakis12a}. Changing
back to the original $x_i$ coordinates the effective medium equation
\cite{antonakakis12a}  reads,
\beq
T_{ij}\frac{\partial^2 f_0}{\partial x_i\partial x_j}-\frac{(\Omega^2-\Omega_0^2)}{l^2}f_0=0.
\label{eq:f_0}
\eeq
$T_{ij}$'s are integrated quantities of the leading order and
first order short scale solutions with $T_{ij}=0$ for $i\neq j$ in the present illustrations. As in \cite{antonakakis12a},
assuming Bloch waves, the asymptotic dispersion relation for $\Omega$ reads,
\beq
\Omega\sim\Omega_0-\frac{T_{ij}}{2\Omega_0}\kappa_i\kappa_j,
\label{eq:Omega}
\eeq
where $\kappa_i=K_i-d_i$ and $d_i=0,-\pi/2,\pi/2$ depending on the
band edge in the Brillouin zone about which the asymptotic expansion
originates, and $\Omega_0$ is the standing wave frequency at the
Brillouin zone edge ; these asymptotics give the dashed curves in Fig. \ref{fig2}. For the case of multiple modes originating from the same point, as for example the Dirac-like cones, Eq. (\ref{eq:Omega}) is no longer valid and one obtains three coupled equations for $f_0^{(i)}$ that simplify to three similar isotropic equations, like (\ref{eq:f_0}) with $T_{11}=T_{22}$. 
Note that Eq. (\ref{eq:f_0}) is of second order and not fourth. The
fourth order plate equation yields two types of solutions namely
propagating and exponentially decaying. Indeed in the long scale,
solutions follow a second order PDE and the remaining information of
the fourth order problem is enclosed in the $T_{ij}$ integrated
quantities. This homogenization theory is not limited to long-waves
relative to the microstructure, one apparent failing is that the
asymptotics appear to be only valid near the band edge frequencies but
further refinements are possible, using foldings of the Brillouin
zone, that extend the theory to provide complete coverage of the
dispersion curves and provide accuracy at all frequencies.

We give three potential applications of the
platonic crystal described by Fig. \ref{fig2}. A finite array of the
PC is embedded within an infinite elastic plate and the full finite
element simulations (making use of specially designed perfectly
matched layers \cite{farhat11} are compared to solutions
constructed using the continuum long-scale HFH theory to replace the
finite array; standard continuity conditions then join this HFH
material to the surrounding infinite elastic plate. This theory captures a rich array of behaviour, for instance
the material displays strong anisotropy with propagation along
characteristics in Fig. \ref{fig1}(b) that is well captured by HFH in
Fig. \ref{fig1}(d); the coefficients
$T_{ij}$ of the HFH theory are of opposite sign ($T_{11}=25.65$ and
$T_{22}=-11.18$) and 
capture the strong anisotropy of the medium and the effective material
becomes hyperbolic rather than elliptic, which unveils the similar
lensing effect without negative refractive index to that observed in \cite{luo02a}. Also shown in
Fig. \ref{fig1}(a) are the shielding effect of the zero frequency stop
band and strong guiding of energy within a PC slab Fig. \ref{fig1}(c).

 Using the vanishing group velocity of the second
dispersion curve in Fig. \ref{fig2}, in
the neighborhood of $\Gamma$, a strongly directive antenna is created as
shown in Fig. \ref{fig3}(a,b) and (e). The effective medium for the directive antennas is governed by Eq. (\ref{eq:f_0}) with $T_{11}=T_{22}=6.2524$.
%In Fig.  \ref{fig3}(e) the group velocity is calculated by differentiating Eq. (\ref{eq:Omega}) with respect to $\kappa_1$ to obtain,
%\beq
%\Omega_{,\kappa_1}\sim-\frac{T_{11}}{\Omega_0}\kappa_1,
%\label{eq:GroupVels}
%\eeq
%and similarly for $\kappa_2$.
Fig. \ref{fig3}(c) and (d) shows an endoscope effect
wherein a point source located
close to a tilted array leads
to focussing and a strong localised beam. The behaviour near point $M$ of the Brillouin zone is responsible
for the focussing effects of Fig. \ref{fig1}(b) and Fig. 
\ref{fig3}(c).
% Even for symmetric cells, where the irreducible
%Brillouin zone is presented in the inset to fig. \ref{fig2}, 
 Here the closest standing wave frequency is at 
point $M$ and unequal $T_{ii}$ coefficients lead to very strong anisotropy within the effective
material. By considering only a portion
of the Brillouin zone we must bear in mind that this asymptotic
solution is also valid at $N$ and there is a second $f$ equation with
the $T_{11}$ and $T_{22}$ interchanged that is used. 
 %This counterintuitive result is supported by the fact that all
 %frequencies of point $M$ are also present for point $N(0,\pi/2)$
 %where the effective medium equation will be the same as for point
 %$M$ but with $T_{11}$ and $T_{22}$ reversed so that an expected
 %symmetry in the equations is restated. Nevertheless these two
 %equations represent two separate anisotropic media and the effects
 %in figures \ref{fig1} and \ref{fig4}(c) encorporate both effective
 %media by superposition. Equation (\ref{eq:f_0}) for the first mode
 %near point $M$ reads,
The asymptotic PDE for HFH is Eq. (\ref{eq:f_0}) with $T_{11}=-9.649$ and $T_{22}=-4.935$.
%\begin{equation}
%\begin{array}{cc}
%&-9.6497f_{0,X_1X_1}+4.7149f_{0,X_2X_2} \nonumber \\
%&+(-9.6497\kappa_1^2+4.7149\kappa_2^2)f_0=0.
%\end{array}
%\label{eq:Meq}
%\end{equation}
$\Omega_2$ is determined using the the frequencies
stated in Figs \ref{fig1}(b), \ref{fig3}(c) and the standard expansion for $\Omega$. 

Last, but not least, the triple crossings in the dispersion diagram
in Fig. \ref{fig2} comprise 
Dirac-like cones with a flat mode passing through
the vertex. This is highly interesting given that Dirac-like cones are
normally limited to graphene-like hexagonal structures \cite{neto09a}; the
current situation involves a square lattice is akin to the Dirac-like cones
for photonic crystals 
recently described in \cite{liu11a,huang11a}. 
 One can use the properties of Dirac cones-like to reduce the
scattering of a clamped obstacle
in a PC, as demonstrated
in Fig. \ref{fig4}(a),(b). This behaviour is also captured by HFH
asymptotics: 
In this particular case three coupled equations emerge with variables $f_0^{(i)}$ for $i=1,2,3$. 
The system decouples to yield the same governing equation for all
three functions $f_0^{(i)}$ as,
\beq
8\pi^6\frac{\partial^2 f_{0}}{\partial x_i^2}+\frac{(\Omega^2-\Omega_0^2)^2}{l^2}f_0=0,
\label{eq:f_0Dirac}
\eeq
where the coefficient in front of $f_0$ comes from first order correction $\Omega_1$ and the change into the original coordinates; the numerics using HFH are shown in Fig. \ref{fig4}(c). The front and back scattering do not depend on the shape of the obstacle but mostly on its area. Fig. \ref{fig5}(c) and (d) show the scattering of three different obstacles (a square, a tilted square and a circle) are virtually indistinguishable, whereas in panels (a) and (b) the three curves can be clearly told apart. The platonic crystal is used to destroy
any one-to-one correspondence between the scattered field and the shape of an obstacle, which is the essence of cloaking in
impedance tomography \cite{greenleaf03,kohn08}. However, we note that the present cloaking is constrained to normal incidence
and is reminiscent of \cite{smith10} which is a more elaborate type of cloaking somewhat constrained to the eikonal limit of transformation optics with photonic band gap media \cite{liang11}.

In conclusion, a simplified model of elasticity via the thin-plate
equation allows for analytical and numerical studies opening
interesting possibilities in the design of flat lens, directive
antenna, endoscope, shielding and a cloak for flexural waves.
Moreover, such designs could be scaled up in order to achieve some
control of seismic surface waves \cite{menard12}, \cite{kim12}, indeed this work is
motivated by a requirement at CERN to control elastic waves created by
thermal shock. We also show, for the first time, that the computations
using the recently developed HFH are capable of capturing the
fundamental wave propagation features of these various possibilities,
including counter-intuitive physics of Dirac-like cones \cite{liu11a}.

R.V.C. thanks the EPSRC (UK) for support through research grant nunmber EP/J009636/1.
S.G. is thankful for an ERC starting grant (ANAMORPHISM).

\bibliographystyle{apsrmp4-1}
\bibliography{arxiv2013_v2}

\end{document}